# Photonuclear reactions on $^{59}$Co at bremsstrahlung end-point energies of 40–130 MeV

F.A. Rasulova[a,b,*], S.S. Belyshev[c,d], M.A. Demichev[a], D.L. Demin[a], S.A. Evseev[a], N.Yu. Fursova[c,d], M.I. Gostkin[a], J.H. Khushvaktov[a,b], V.V. Kobets[a], A.A. Kuznetsov[c,d], S.V. Rozov[a], E.T. Ruziev[b], A.A. Solnyshkin[a], T.N. Tran[a,e], E.A. Yakushev[a], B.S. Yuldashev[a,b]

[a] Joint Institute for Nuclear Research, Dubna, Russia
[b] Institute of Nuclear Physics of the Academy of Sciences of the Republic of Uzbekistan, Tashkent, Uzbekistan
[c] Skobeltsyn Institute of Nuclear Physics of Lomonosov Moscow State University, Moscow, Russia
[d] Faculty of Physics of Lomonosov Moscow State University, Moscow, Russia
[e] Institute of physics, Vietnam Academy of Science and Technology, Hanoi, Vietnam

[*]*rasulova@jinr.ru*

**Abstract**

Relative yields have been measured in the 40-130 MeV bremsstrahlung induced reactions of $^{59}$Co. The experiments have been performed with the beam from the electron linear accelerator LINAC-200 using the activation and off-line γ-ray spectrometric technique. The bremsstrahlung photon flux has been calculated with the Geant4 program. The cross sections were calculated by using computer code TALYS-1.96 with different models and are found to be in good agreement with the experimental data.

## 1. Introduction

The photon induced reaction cross sections of $^{59}$Co are important for designing conventional reactors, fast reactors, and accelerator-driven systems (ADSs). The high energy photon induced reactions of $^{59}$Co can be used for production of important medical isotopes, such as $^{55}$Co, $^{56}$Co, $^{57}$Co, and $^{58}$Co, due to their suitable decay characteristics. Also, $^{59}$Co is often used as a monitor target during the study of photonuclear reactions to determine the photon of gamma rays or the accelerator current.

The $^{59}$Co($\gamma$,$x$n)$^{58-56}$Co reaction cross sections already covered in Refs. [1–5] are only for monoenergetic photons in the energy range of 9.65 to 36.504 MeV. The flux-averaged cross sections for the $^{59}$Co($\gamma$,$x$n)$^{58-55}$Co reaction have been measured using the method of activation at the bremsstrahlung end-point energies of 15 MeV [6], 22 MeV [7], 25 MeV [8], 30 MeV [9], 65, and 75 MeV [10]. In Ref. [11], the yields of the reactions ($\gamma$, 2$n$), ($\gamma$, 3$n$), ($\gamma$, 4$n$), and $^{59}$Co($\gamma$, 1$n$2$p$)$^{56}$Mn, relative to the yield of the reaction (γ, n), were determined using the measurements of gamma-ray activities induced when irradiating $^{59}$Co with the 35- and 54-MeV bremsstrahlung. In Ref. [12], the $^{59}$Co(γ, 1n2p)$^{56}$Mn reaction was investigated in the energy range of 30–260 MeV, and differential effective cross sections were calculated from yield curves using the Photon Difference Method. In Refs. [13,14], the data were presented for photospallation yields in units of mb per equivalent photon from a $^{59}$Co target measured at the bremsstrahlung end-point energies of 30 to 1050 MeV.

The present work is concerned with obtaining the new nuclear data for bremsstrahlung-induced photonuclear reactions on $^{59}$Co in the energy range of 40–130 MeV and comparing these data with the literature ones [7–10] and theoretical values computed with the program package TALYS1.96 [15].

## 2. Materials and methods

The experiment was carried out with electron beams at the energies of 40–130 MeV. A tungsten plate with a size of 4.5 × 4.5 cm and a thickness of 0.5 cm was used as a converter. In the experiments ($E_e$ = 60, 80, 105, and 130 MeV), cobalt metal samples were irradiated with a bremsstrahlung flux formed in the tungsten converter. For comparing the results of experiments performed under different conditions at an energy of 40 MeV, the target was directly irradiated with the electron beam. The main parameters of the experiments are given in Table 1.

Table 1. Main parameters of the experiments

|  | Experiment 1 | Experiment 2 | Experiment 3 | Experiment 4 | Experiment 5 |
|---|---|---|---|---|---|
| Energy of electrons, MeV | 40 | 60 | 80 | 105 | 130 |
| Electron beam pulse current, mA | 48 | 40 | 40 | 58 | 50 |
| Irradiation time, min | 20 | 25 | 15 | 20 | 15.5 |
| Integral number of electrons incident on the tungsten converter, ×10$^{15}$ | 7.2 ± 0.72 | 7.5 ± 0.75 | 4.5 ± 0.45 | 8.7 ± 0.87 | 5.8 ± 0.58 |
| Mass of cobalt target, mg | 312.0 | 155.7 | 145.8 | 729.5 | 372.8 |
| Dimensions of cobalt target, cm×cm | 1.25×1.25 | 1.35×0.6 | 1.25×0.68 | 1.25×1.25 | 1.25×1.25 |
| Cooling time, min | 50 | 56 | 56 | 39 | 107 |
| Measuring time of spectra, day | 21.8 | 12.4 | 8.2 | 8.1 | 28.8 |

The electron linac was operated in the stable mode with a beam pulse width of 2 μs and a repetition rate of 10 Hz. The electron current was measured with a high sensitivity inductive current sensor [16]. After irradiation, the target was transferred to the test room where the induced activity in the irradiated target was measured using a high-purity germanium detector (HPGe detector). The spectra of γ-quanta of irradiated targets in the energy range from 50 keV to 3.7 MeV were measured using a CANBERRA GC3018 HPGe detector with a volume of 145 cm$^3$. We used the HPGe γ-detector with resolutions of 0.8 keV at 122 keV and of 1.76 keV at 1332 keV in combination with standard measurement electronics and a 8K ADC/MCA (Multiport II Multichannel Analyzer). The relative efficiency of the detector is 34.7%. For the γ-activity measurements, the activated sample were placed at 56 mm and 130 mm from the detector surface to maintain the detector dead time at approximately 5% and to reduce counting losses due to the coincidence summing effect. Compton background has been missed about 3 hours after irradiation, so the targets were located 10 mm from the detector surface for measuring γ-rays of long-lived isotopes in the target.

Energy and efficiency calibrations of the HPGe detector were carried out using the standard gamma-ray sources $^{22}$Na, $^{54}$Mn, $^{57}$Co, $^{65}$Zn, $^{88}$Y, $^{113}$Sn, $^{133}$Ba, $^{139}$Ce, $^{152}$Eu, $^{207}$Bi, and $^{241}$Am. The gamma-ray spectra were processed using the program DEIMOS32 [17] which fits the count area of the full-energy peaks with the Gaussian function. The processed peaks were identified on the basis of gamma-ray energy and intensity and the half-life of generated residual nuclei. The typical γ-ray spectra of reaction products from irradiated cobalt with the bremsstrahlung end-point energy of 130 MeV are shown in Fig. 1.

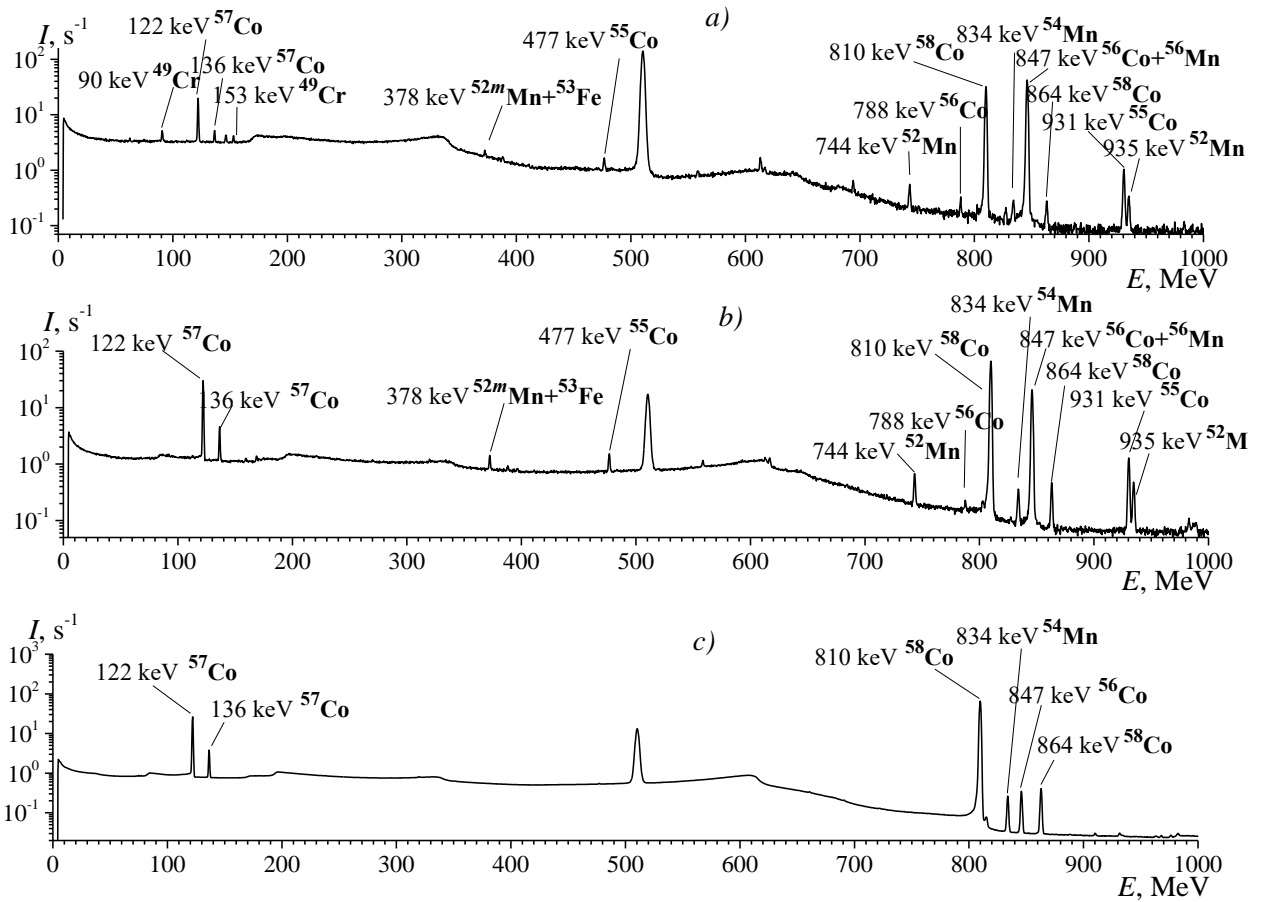

Figure 1. Spectra of residual activity of the irradiated sample of $^{59}$Co (top-to-bottom) 2 h (*a*), 6 h (*b*), and 40 days (*c*) after irradiation. The spectra measurement duration was 10 min (*a*), 1 h (*b*), and 7 days (*c*), respectively. The bremsstrahlung end-point energy used for the irradiation was 130 MeV

The experimental yields of the reactions $Y_{exp}$ were normalized to one electron of the accelerated beam incident on the bremsstrahlung target and calculated using the following formula:

$$Y_{exp} = \frac{S_p \cdot C_{abs}}{\varepsilon \cdot I_\gamma} \frac{t_{real}}{t_{live}} \frac{1}{N} \frac{1}{N_e} \frac{e^{\lambda \cdot t_{cool}}}{(1 - e^{-\lambda \cdot t_{real}})} \frac{\lambda \cdot t_{irr}}{(1 - e^{-\lambda \cdot t_{irr}})}, \qquad (1)$$

where $S_p$ is the full-energy-peak area; $\varepsilon$ is the full-energy-peak detector efficiency; $I_\gamma$ is the gamma emission probability; $C_{abs}$ is the correction for self-absorption of gamma rays in the sample; $t_{real}$ and $t_{live}$ are the real time and live time of the measurement, respectively; $N$ is the number of atoms in the activation sample; $N_e$ is the integral number of incident electrons; $\lambda$ is the decay constant; $t_{cool}$ is the cooling time; and $t_{irr}$ is the irradiation time. The main γ-ray energies and intensities used to determine the yield of reaction products are given in Table 2. The nuclear data given in Table 2 are taken from the ref. [18].

Table 2. Spectroscopic data from ref. [18] for product-nuclei from photonuclear reactions

| Product of reaction | Half-life | Reaction | Threshold, MeV | $E_\gamma$, keV ($I_\gamma$, %) |
|---|---|---|---|---|
| $^{58g}$Co | 70.86 d | $^{59}$Co(γ, n) | 10.45 | 810.76 (99.45) |
| $^{58m}$Co | 9.10 h | $^{59}$Co(γ, n) | 10.48 | 24.89 (0.0397) |
| $^{57}$Co | 271.74 d | $^{59}$Co(γ, 2n) | 19.03 | 122.06 (85.6), 136.47 (10.68) |
| $^{56}$Co | 77.236 d | $^{59}$Co(γ, 3n) | 30.40 | 846.77 (99.94), 1037.84 (14.05), 1238.29 (66.46), 1771.36 (15.41), 2598.5 (16.97) |
| $^{55}$Co | 17.53 h | $^{59}$Co(γ, 4n) | 40.49 | 477.2 (20.2), 931.1 (75), 1408.5 (16.9) |
| $^{53}$Fe | 8.51 m | $^{59}$Co(γ, 5n1p) | 58.93 | 377.88 (42) |
| $^{52}$Fe | 8.275 h | $^{59}$Co(γ, 6n1p) | 69.61 | 168.688 (99.2) |
| $^{56}$Mn | 2.58 h | $^{59}$Co(γ, 1n2p) | 27.98 | 846.76 (98.85), 1810.73 (26.9), 2113.09 (14.2) |
| $^{54}$Mn | 312.2 d | $^{59}$Co(γ, 1n1α) $^{59}$Co(γ, 3n2p) | 17.17 45.46 | 834.85 (99.98) |
| $^{52g}$Mn | 5.59 d | $^{59}$Co(γ, 3n1α) $^{59}$Co(γ, 5n2p) | 38.16 66.45 | 744.23 (90), 935.54 (94.5), 1434.06 (100) |
| $^{52m}$Mn | 21.1 min | $^{59}$Co(γ, 3n1α) $^{59}$Co(γ, 5n2p) | 38.54 66.83 | 377.74 (1.68), 1434.06 (98.2) |
| $^{51}$Cr | 27.7 d | $^{59}$Co(γ, 3n1p1α) $^{59}$Co(γ, 5n3p) | 44.71 73.00 | 320.08 (9.91) |
| $^{49}$Cr | 42.3 min | $^{59}$Co(γ, 5n1p1α) $^{59}$Co(γ, 7n3p) | 66.97 95.26 | 62.29 (16.4), 90.64 (53.2), 152.93 (30.3) |

The yields $Y_{theor}$ of photonuclear reactions representing the convolution of the photonuclear reactions cross section σ(E) and the distribution density of the number of bremsstrahlung photons over energy per one electron of the accelerator $W(E, E_{\gamma max})$ were determined as a result of the experiment. For the yield measurement of a natural mixture of isotopes, the result is the yield of isotope production in all possible reactions on the natural mixture:

$$Y_{theor} = \int_{E_{th}}^{E_{\gamma max}} \sigma(E) W(E, E_{\gamma max}) dE \qquad (2)$$

where $E_{\gamma max}$ is the kinetic energy of electrons hitting the tungsten radiator, $E$ is the energy of bremsstrahlung photons produced on the radiator, $E_{th}$ is the threshold of the studied photonuclear reaction.

The use of the relative yields makes it possible to obtain the dependence of the probability of photonuclear reactions on the maximum energy of bremsstrahlung under different experimental conditions. The calibration with respect to the yield of the most probable reaction excludes the influence of the total photon absorption cross section. In our case the $^{59}$Co$(\gamma,n)^{58m+g}$Co reaction was chosen as a dominant reaction because of its wide application in this capacity. The half-lives of the isotopes $^{58m}$Co and $^{58g}$Co are 9.10 hours and 70.86 days, respectively. $^{58m}$Co decays by IT with branching fraction of 100%. Since $^{58m}$Co has the single gamma line with $E_\gamma = 24.89$ keV, it is not possible to experimentally determine the yield of this isotope by the usual method from the peak in the residual activity spectrum; the detector used was capable of detecting γ-quanta starting from 50 keV. We can determine only the cumulative yield of $^{58m+g}$Co. Thus, the net photopeak counts of $^{58g}$Co from different spectra were taken after the decay of more than ten half-lives of $^{58m}$Co.

Theoretical values of the relative yields can be calculated using the following formula:

$$Y_{rel,i} = \frac{\int_{E_{th}}^{E_{\gamma max}} \sigma_i(E) W(E, E_{\gamma max}) dE}{\int_{E_{th}}^{E_{\gamma max}} \sigma_{(\gamma,n)}(E) W(E, E_{\gamma max}) dE} \qquad (3)$$

Owing to the assumption on the unchanged shape of the bremsstrahlung spectrum, the bremsstrahlung photon production cross section $\sigma(E, E_{\gamma max})$ should be taken as the function $W(E, E_{\gamma max})$:

$$Y_{rel,i} = \frac{\int_{E_{th}}^{E_{\gamma max}} \sigma_i(E) \sigma(E, E_{\gamma max}) dE}{\int_{E_{th}}^{E_{\gamma max}} \sigma_{(\gamma,n)}(E) \sigma(E, E_{\gamma max}) dE} \qquad (4)$$

where $\sigma(E, E_{\gamma max})$ is calculated based on the Zeltzer-Berger tables [19].

The total and partial cross sections $\sigma(E)$ of the $^{59}$Co$(\gamma; xnyp)$ reaction were computed for monochromatic photons with the TALYS code. The phenomenological models that we will use below are generally parameterized in terms of Lorentzian forms with giant resonance parameters:
- $\sigma_{Xl}$: strength of the giant resonance,
- $E_{Xl}$: energy of the giant resonance,
- $\Gamma_{Xl}$: width of the giant resonance.

The photoreaction cross section data have been compared to theoretical calculations on the basis of three standard models of the $\gamma$SF, namely Kopecky-Uhl generalized Lorentzian model (GLO) [20], Brink-Axel standard Lorentzian model (SLO) [21,22], Simplified Modified Lorentzian model (MLO) [23], also in addition to phenomenological models (GLO, SLO and MLO) two microscopic models Skyrme-Hartree-Fock-Bogoliubov model with QRPA by Goriely et al. (HFB+QRPA) [24] and Gogny-Hartree-Fock-Bogoliubov model with QRPA by Goriely et al. (GHF+QRPA) [25] are used. This approach is quite natural, since it involves the use of a single-particle nuclear level scheme and, therefore, takes into account the individual properties of each nucleus. Fig. 2 shows experimental data and results of TALYS models on the cross section of reactions $(\gamma, n)$, $(\gamma, 2n)$, $(\gamma, n)+(\gamma, np)$ and $(\gamma, 1n2p)$ on $^{59}$Co.

GLO was suggested to predict the correct behavior of the E1 γ-strength functions in the low-energy range by adopting the temperature dependent width along with the non-zero limiting value as $E_\gamma$ goes to zero. We use the SLO for all multipoles higher than 1. The MLO model was developed to provide an improved estimate of the E1 and M1 PSFs for all nuclei. The HFB+QRPA model is based on the Skyrme force where some phenomenological corrections are introduced to take the damping of the collective motion as well as the deformation effects into account. The GHF+QRPA model allows for a consistent description of axially symmetric deformations and includes phenomenologically the impact of multiparticle multihole configuration as a function of their densities.

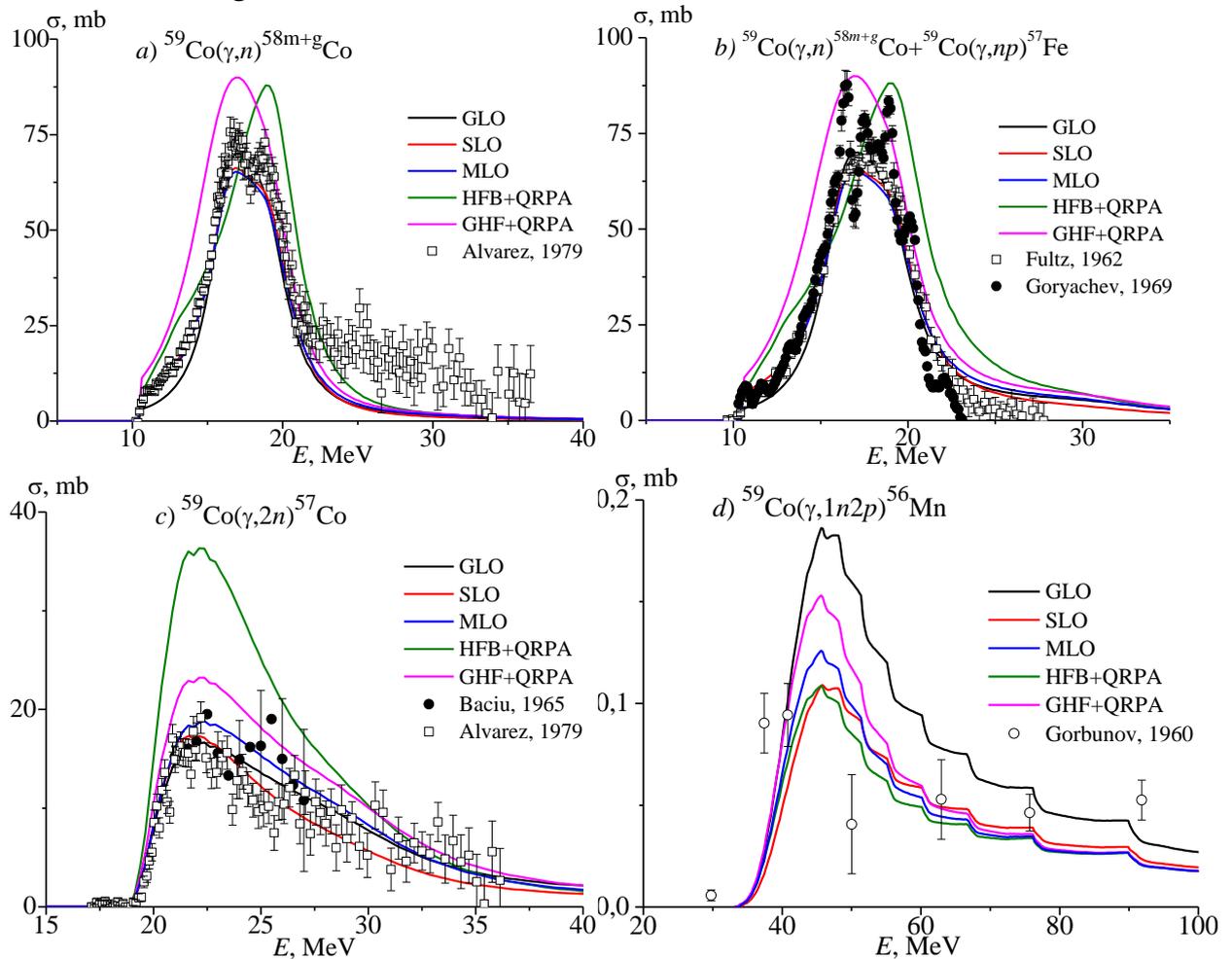

Fig. 2. Experimental data and results of TALYS models on the cross section of reactions $(\gamma, n)$ [3], $(\gamma, 2n)$ [3,7], $(\gamma, n)+(\gamma, np)$ [4,9] and $(\gamma, 1n2p)$ [12] on $^{59}$Co

The photon flux in the sample was simulated using Geant4 [26]. The energy spectrum of bremsstrahlung in the location of cobalt samples is shown in Fig. 3. In the experiments at 60 and 80 MeV, the shape of the electron beam was round with a diameter 5.5 mm. In the experiments at 40, 105 and 130 MeV, the shape of the electron beam was rectangle with a size 2×20 mm.

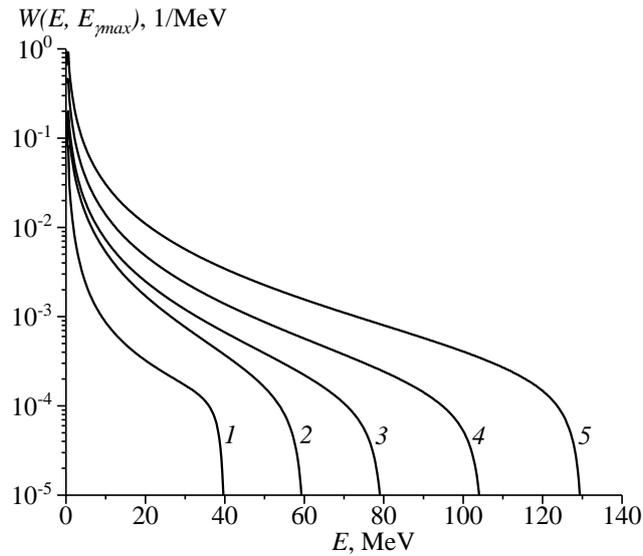

Figure 3. Distribution density of the number of bremsstrahlung photons at the energies of 40 (*1*), 60 (*2*), 80 (*3*), 105 (*4*), and 130 MeV (*5*)

## 3. Results and discussion

The relative yields $Y_{rel}$ of the photonuclear reactions measured in the bremsstrahlung end-point energy range of 40–130 MeV from the present experiment are given in Table 3 and in Figs. 4–7.

*Reactions with neutron emissions*

Fig. 4 shows relative yields of the $^{59}Co(\gamma, xn)^{57,56,55}Co$ reactions. Within the limits of error our result on relative yields of the $^{59}Co(\gamma,2n)^{57}Co$ reaction in 40 MeV agrees well with calculation based on models GLO, SLO and MLO, and with literature data [13]. The experimental points at 60, 67 [13] and 80 MeV are lower than the calculated data. Our result on relative yield of the $^{59}Co(\gamma,3n)^{56}Co$ reaction in 40 MeV agrees well with calculation based on models SLO. The experimental points at 40 [13], 60, 67 [13] and 80 MeV are lower than the calculated data. The experimental points on relative yield of the $^{59}Co(\gamma,4n)^{55}Co$ reaction are lower than the calculated data. Since there are not sufficient experimental points in this area, we cannot draw clear conclusions on the cross sections.

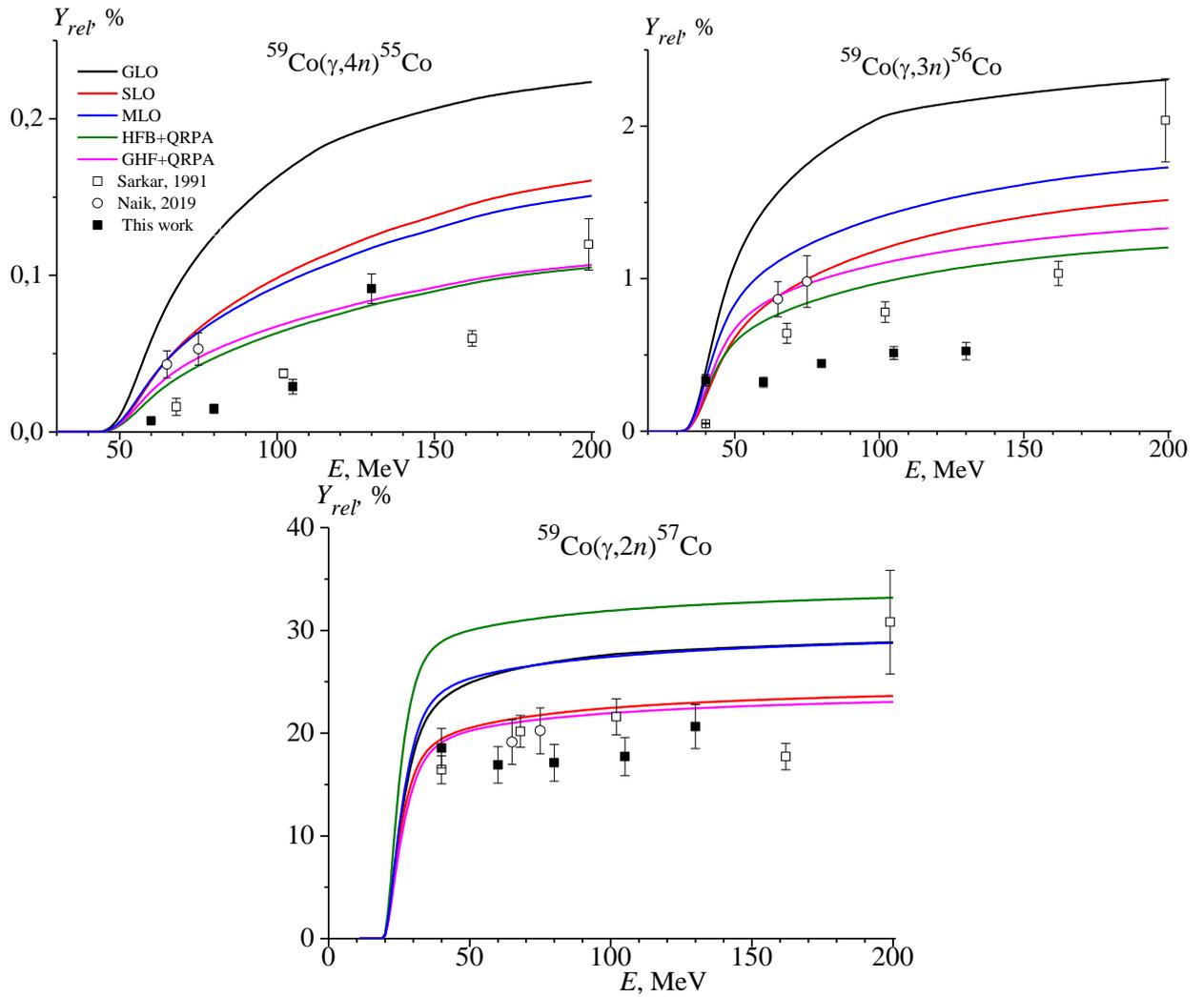

Figure 4. Relative yields of the $^{59}$Co($\gamma$, x$n$)$^{57,56,55}$Co reactions as a function of the bremsstrahlung end-point energy from the present work (solid squares) and literature data [4] (open triangles), [10] (open circles), [13] (open squares) and the theoretically calculated using different models for $E$1 gamma-ray strength function in TALYS values

*Reactions with the emission of one proton and several neutrons*

It can be seen from Fig. 5 that the experimentally obtained relative yields for the $^{59}$Co($\gamma$,6$n$1$p$)$^{52}$Fe reaction at energies 162 and 200 MeV on the basis of the literature data [13] are not in agreement with the values simulated with TALYS, according to different models. Experimental point in 130 MeV are in agreement with the calculated values based on different models without GLO.

It can be seen from Fig. 5 that the experimentally obtained relative yields for the $^{59}$Co($\gamma$,5$n$1$p$)$^{53}$Fe reaction at energy 102 MeV [13] is not in agreement with the values simulated with TALYS, according to different models. Experimental point in 105 MeV is in agreement with the calculated values based on model HFB+QRPA.

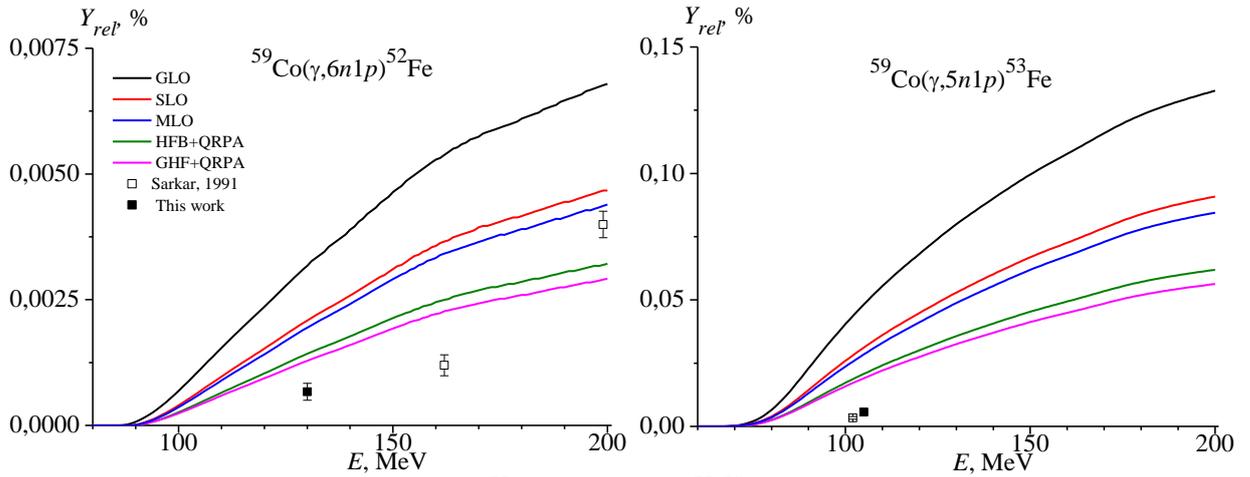

Figure 5. Relative yields of the $^{59}$Co($\gamma$, x$n$1$p$)$^{52,53}$Fe reactions as a function of the bremsstrahlung end-point energy from the present work (solid squares) and literature data [4] (open triangles), [10] (open circles), [13] (open squares) and the theoretically calculated using different models for $E$1 gamma-ray strength function in TALYS values

*Reactions with the emission of two protons and several neutrons*

It can be seen from Fig. 6 that the experimentally obtained relative yields for the production of $^{52g}$Mn at the bremsstrahlung end-point energies of 60, 80 and 105 MeV are in agreement with the values simulated with TALYS, according to different models (SLO and MLO). It can be seen from Fig. 6 that the experimentally obtained relative yields for the production of $^{52m}$Mn at the bremsstrahlung end-point energies of 60, 80 and 105 MeV are in agreement with the values simulated with TALYS, according to model SLO and MLO. Literature data [13] lie much lower than the calculated curves.

It can be seen from Fig. 6 that the experimentally obtained relative yields for the production of $^{54}$Mn at the bremsstrahlung end-point energy of 40 MeV is in agreement with the values simulated with TALYS, according to model GLO. It can be seen from the Fig. 6 that the relative yields for the production of $^{54}$Mn are an order of magnitude greater than the relative yields for the reaction $^{59}$Co($\gamma$, 1$n$2$p$)$^{56}$Mn.

It can be seen from Fig. 6 that the experimentally obtained relative yields for the $^{59}$Co($\gamma$,1$n$2$p$)$^{56}$Mn reaction at the bremsstrahlung end-point energies of 40–100 MeV are not in agreement with the values simulated with TALYS.

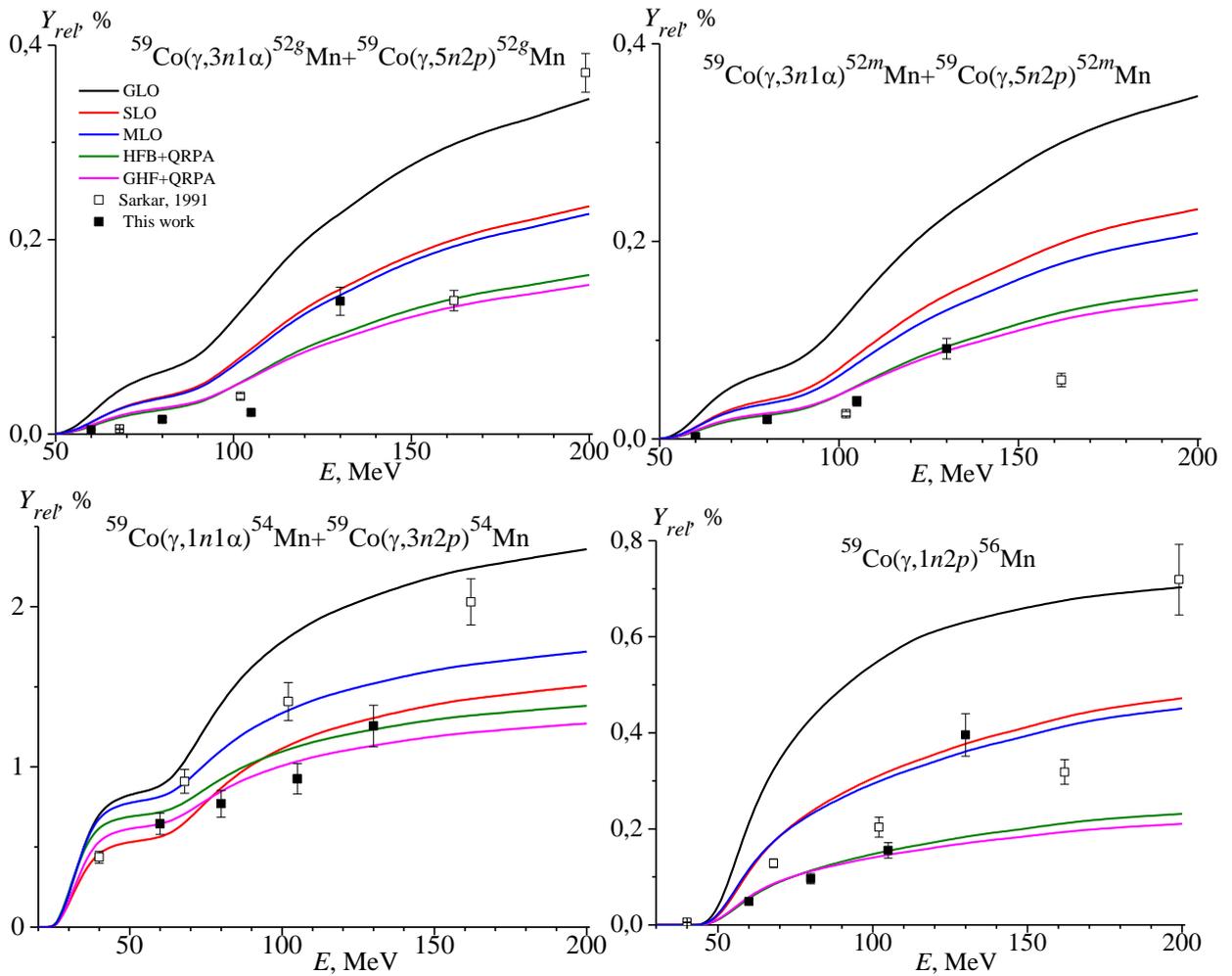

Figure 6. Relative yields for the production of $^{52g,52m,54,56}$Mn from reactions $^{59}$Co($\gamma$, x$n$1$\alpha$) and $^{59}$Co($\gamma$,x$n$2$p$) as a function of the bremsstrahlung end-point energy from the present work (solid squares), literature data [13] (open squares) and the theoretically calculated using different models for $E$1 gamma-ray strength function in TALYS values

From Fig. 6, it is clear that relative yields for reactions producing $^{54}$Mn, $^{52g}$Mn and $^{52m}$Mn are not constantly increasing functions. This is due to the fact that in the initial energy region the main channels for the formation of the $^{54}$Mn, $^{52g}$Mn and $^{52m}$Mn isotopes are reactions with the emission of an alpha-particle. Near the energies of reaction threshold, the $^{59}$Co($\gamma$, x$n$2$p$)$^{54,52g,52m}$Mn reaction is accompanied with the release of alpha-particles and thus it cross-sections decrease. A further increase in the relative yields is associated with the reactions $^{59}$Co($\gamma$, x$n$2$p$)$^{54,52g,52m}$Mn.

*Reactions with the emission of three protons and several neutrons*

It can be seen from Fig. 7 that the experimentally obtained relative yields for the production of $^{49}$Cr at energies 102 from literature [13] is not in agreement with the values simulated with TALYS based on different models. However, our experimental points are in agreement with the simulated values based on TALYS with models MLO and HFB+QRPA. It can be seen from Fig. 7 that the experimentally obtained relative yields for the production of $^{51}$Cr at energies 102 and 162 MeV from literature [13] are not in agreement with simulated values based on TALYS with different models.

As can be seen from Fig. 7 that the relative yields for reactions producing $^{51}$Cr and $^{49}$Cr are not constantly increasing functions. This is due to the fact that in the initial energy region the main channels for the formation of $^{51}$Cr and $^{49}$Cr isotopes are reactions with the emission of an alpha-particles. At energies close to the threshold of the reactions $^{59}$Co($\gamma$, 5n3p)$^{51}$Cr and $^{59}$Co($\gamma$, 7n3p)$^{49}$Cr, the cross sections for reactions with the emission of alpha-particles decrease, therefore, in this energy region in Fig. 7, a decrease in values is observed.

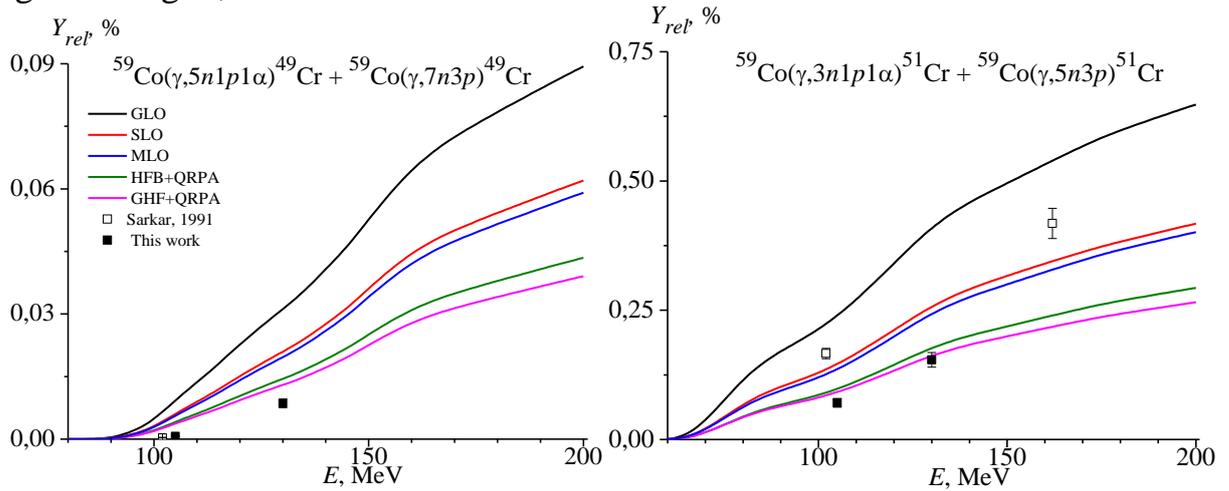

Figure 7. Relative yields for the production of $^{49,51}$Cr from reactions $^{59}$Co($\gamma$, xn1p1α) and $^{59}$Co($\gamma$,xn3p) as a function of the bremsstrahlung end-point energy from the present work (solid squares), literature data [13] (open squares) and the theoretically calculated using different models for $E$1 gamma-ray strength function in TALYS values

Table 3. Relative yields of the photonuclear reactions on $^{59}$Co at the bremsstrahlung end-point energies of 40–130 MeV from the present experiment

| Isotope | Bremsstrahlung end-point energy. MeV | Relative yields. % | | | | | |
|---|---|---|---|---|---|---|---|
| | | Experiment | TALYS-1.96 | | | | |
| | | | GLO | SLO | MLO | HFB+QRPA | GHF+QRPA |
| $^{57}$Co | 40 | 18 ± 2 | 23 | 19 | 24 | 29 | 19 |
| | 60 | 17 ± 2 | 26 | 21 | 26 | 31 | 21 |
| | 80 | 17 ± 2 | 27 | 22 | 27 | 31 | 21 |
| | 105 | 18 ± 2 | 28 | 23 | 28 | 32 | 22 |
| | 130 | 21 ± 2 | 28 | 23 | 28 | 32 | 22 |
| $^{56}$Co | 40 | 0.33 ± 0.04 | 0.40 | 0.23 | 0.34 | 0.25 | 0.27 |
| | 60 | 0.32 ± 0.03 | 1.45 | 0.81 | 1.04 | 0.72 | 0.84 |
| | 80 | 0.44 ± 0.02 | 1.82 | 1.04 | 1.26 | 0.87 | 1.00 |
| | 105 | 0.51 ± 0.04 | 2.08 | 1.22 | 1.43 | 0.99 | 1.12 |
| | 130 | 0.52 ± 0.05 | 2.17 | 1.34 | 1.55 | 1.08 | 1.20 |
| $^{55}$Co | 60 | 0.0071 ± 0.0008 | 0.059 | 0.033 | 0.034 | 0.022 | 0.026 |
| | 80 | 0.015 ± 0.003 | 0.125 | 0.074 | 0.071 | 0.047 | 0.052 |
| | 105 | 0.029 ± 0.005 | 0.170 | 0.103 | 0.098 | 0.067 | 0.071 |
| | 130 | 0.091 ± 0.009 | 0.195 | 0.125 | 0.117 | 0.081 | 0.084 |
| $^{53}$Fe | 105 | 0.0056 ± 0.0006 | 0.048 | 0.031 | 0.029 | 0.021 | 0.019 |
| $^{52}$Fe | 130 | 0.0006 ± 0.0001 | 0.0032 | 0.0021 | 0.0019 | 0.0014 | 0.0013 |
| $^{56}$Mn | 60 | 0.048 ± 0.005 | 0.21 | 0.11 | 0.12 | 0.05 | 0.06 |
| | 80 | 0.096 ± 0.011 | 0.43 | 0.24 | 0.23 | 0.11 | 0.11 |
| | 105 | 0.155 ± 0.016 | 0.56 | 0.32 | 0.31 | 0.15 | 0.15 |
| | 130 | 0.396 ± 0.044 | 0.63 | 0.38 | 0.36 | 0.18 | 0.17 |
| $^{54}$Mn | 60 | 0.65 ± 0.07 | 0.88 | 0.56 | 0.81 | 0.72 | 0.64 |
| | 80 | 0.77 ± 0.08 | 1.39 | 0.87 | 1.10 | 0.92 | 0.85 |
| | 105 | 0.93 ± 0.10 | 1.85 | 1.16 | 1.38 | 1.13 | 1.03 |
| | 130 | 1.26 ± 0.14 | 2.06 | 1.31 | 1.52 | 1.23 | 1.13 |
| $^{52g}$Mn | 60 | 0.0044 ± 0.006 | 0.021 | 0.012 | 0.012 | 0.008 | 0.009 |
| | 80 | 0.015 ± 0.002 | 0.064 | 0.038 | 0.037 | 0.025 | 0.027 |
| | 105 | 0.022 ± 0.003 | 0.139 | 0.088 | 0.084 | 0.059 | 0.058 |
| | 130 | 0.137 ± 0.017 | 0.227 | 0.149 | 0.143 | 0.103 | 0.098 |
| $^{52m}$Mn | 60 | 0.0025 ± 0.0005 | 0.022 | 0.012 | 0.012 | 0.008 | 0.009 |
| | 80 | 0.019 ± 0.003 | 0.067 | 0.039 | 0.036 | 0.024 | 0.026 |
| | 105 | 0.038 ± 0.005 | 0.138 | 0.085 | 0.076 | 0.054 | 0.053 |
| | 130 | 0.091 ± 0.012 | 0.226 | 0.146 | 0.130 | 0.093 | 0.089 |
| $^{51}$Cr | 105 | 0.071 ± 0.008 | 0.24 | 0.15 | 0.14 | 0.10 | 0.09 |
| | 130 | 0.154 ± 0.014 | 0.41 | 0.26 | 0.24 | 0.18 | 0.16 |
| $^{49}$Cr | 105 | 0.0007 ± 0.0001 | 0.009 | 0.006 | 0.006 | 0.004 | 0.004 |
| | 130 | 0.009 ± 0.001 | 0.031 | 0.021 | 0.020 | 0.014 | 0.013 |

The $E1$ photon absorption by isotopes of the $^{59}$Co ($J^{\pi} = 7/2^-$) nucleus leads to the excitation of the giant dipole resonance. Upon the decay of the giant dipole resonance via a reaction involving the emission of alpha particle and 3 neutrons, there arises a $^{52}$Mn nucleus in the ground or excited state. Cascades of gamma transitions from excited states lead to the production of the isomeric or ground state of $^{52}$Mn. The $^{52m}$Mn isomeric state decays through the electron capture or isomer transition channel. The probability $p$ for isomer transition to the ground state is 1.75%. Usually, the isomeric and ground states strongly differ in spins, therefore, the isomeric ratio is represented in most cases through the yields of high- to low-spin states [28]. In the case of $^{52m,g}$Mn, the spin of the isomeric state of $^{52m}$Mn is $2^+$ (low-spin state, $l$) and that of the unstable ground state of $^{52g}$Mn is $6^+$ (high-spin state, $h$). Therefore, the isomeric ratio of $^{52m,g}$Mn was computed as the ratio of the high- to low-spin states.

The isomeric ratios for the production of $^{52m,g}$Mn from reactions $^{59}$Co($\gamma$, $3n1\alpha$) and $^{59}$Co($\gamma$, $5n2p$) at different bremsstrahlung end-point energies are plotted in Fig. 8. The measured isomeric yield ratios for the pairs $^{52m,g}$Mn at 80, 105 and 130 MeV are $0.79 \pm 0.18$, $0.90 \pm 0.14$ and $1.03 \pm 0.16$, respectively. It can be seen in Fig. 8$a$ that the experimentally obtained isomeric ratio for the pairs $^{52m,g}$Mn at the bremsstrahlung end-point energies of 80–130 MeV are in agreement with the values simulated with TALYS on the basis of GLO and SLO. It can be seen in Fig. 8$b$ that our data on isomeric ratio complement the missing experimental data in the literature [10, 13]. In Fig. 8$b$ the lines to just guide the literature data [10] (blue dashed line) and [13] (red dashed line) are shown. The red line shows exponential growth in the region of 50-600 MeV and further saturation. The blue line coincides with an unchanged linear function surrounded by an isomer ratio value of 0.84. It is necessary to conduct experiments in the energy region of 130-300 MeV to obtain accurate information.

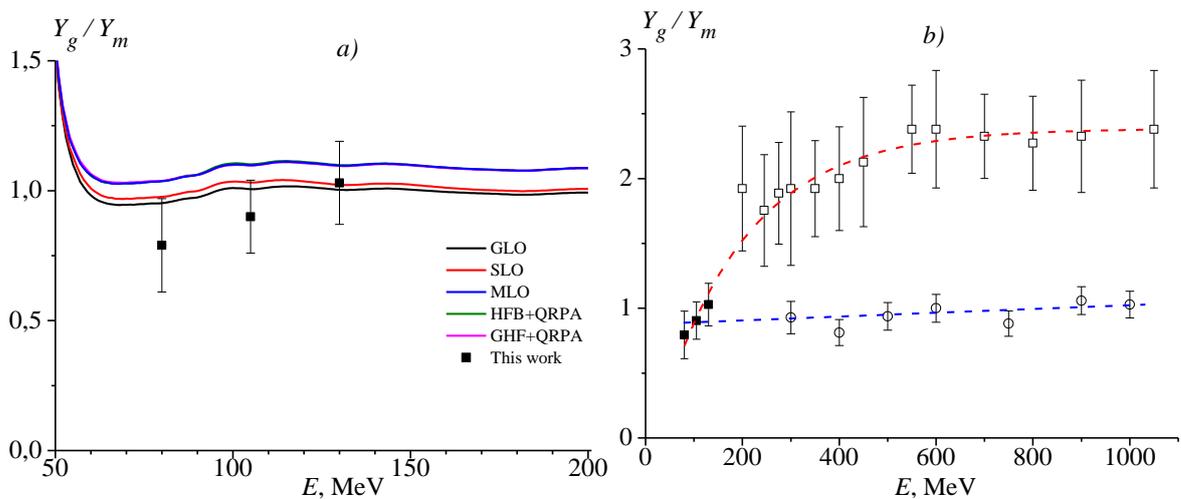

Figure 8. Isomeric yield ratios of the pairs $^{52m,g}$Mn produced in $^{59}$Co($\gamma$, $3n1\alpha$) and $^{59}$Co($\gamma$, $5n2p$) reactions as a function of the bremsstrahlung end-point energy from the present work (solid squares), literature data [10] (open circles), [13] (open

squares) and the theoretically calculated using different models for $E1$ gamma-ray strength function in TALYS values

**4. Conclusion**

Relative yields of the $^{59}$Co($\gamma$,x$n$; $x$=1-4)$^{58-55}$Co, $^{59}$Co($\gamma$, 2$p$x$n$; $x$=1-5)$^{56-52}$Mn, and $^{59}$Co($\gamma$, 3$p$x$n$; $x$=5-7)$^{51,49}$Cr reactions at different bremsstrahlung end-point energies were obtained on the basis of the values simulated with TALYS. The present data along with the literature data are found to be in good agreement with the simulated values. The relative yields for reactions producing $^{54,52g,52m}$Mn and $^{49,51}$Cr are not constantly increasing functions. This is due to the fact that in the initial energy region the main channels for the formation of the $^{54,52g,52m}$Mn and $^{49,51}$Cr isotopes are reactions with the emission of an alpha-particle. Near the energies of reaction threshold, the $^{59}$Co($\gamma$, x$n$2$p$)$^{54,52g,52m}$Mn and $^{59}$Co($\gamma$, x$n$3$p$)$^{49,51}$Cr reactions are accompanied with the release of alpha-particles and thus it cross sections decrease. The experimentally obtained isomeric ratio for the pairs $^{52m,g}$Mn at the bremsstrahlung end-point energies of 80–130 MeV complement the missing experimental data in the literature.

**Acknowledgment**

The authors would like to thank the staff of the linear electron accelerator LINAC-200 (JINR) for their help with the described experiments.
The work was supported by the Ministry of Science and Higher Education of the Russian Federation (grant agreement N◦ 075-15-2021-1360) in part and by the project of the National Center for Physics and Mathematics (NCPM) no. 6 "Nuclear and Radiation Physics," direction 6.5.1.